\documentclass[twocolumn,amssymb,nobibnotes,showpacs,superscriptaddress,aps,reprint]{revtex4-1}
\usepackage{graphicx,amsmath,bm}

\allowdisplaybreaks

\begin{document}

\title{Electromagnetic control of nonclassicality in cavity QED system}

\author{Y. F. Han}
\affiliation{MOE Key Laboratory of Advanced Micro-Structured Materials, School of Physics Science and Engineering, Tongji University, Shanghai 200092, China}
\affiliation{School of Mathematics and Physics, Anhui University of Technology, Ma'anshan 243032, China}

\author{C. J. Zhu}
\email[]{cjzhu@tongji.edu.cn}
\affiliation{MOE Key Laboratory of Advanced Micro-Structured Materials, School of Physics Science and Engineering, Tongji University, Shanghai 200092, China}

\author{X. S. Huang}
\affiliation{School of Mathematics and Physics, Anhui University of Technology, Ma'anshan 243032, China}

\author{Y. P. Yang}
\email[]{yang\_yaping@tongji.edu.cn}
\affiliation{MOE Key Laboratory of Advanced Micro-Structured Materials, School of Physics Science and Engineering, Tongji University, Shanghai 200092, China}



\date{\today}

\begin{abstract}
We present a study of the electromagnetic control of nonclassicality of the outgoing light field in a single atom cavity QED system. By exploring the energy  eigenvalues and eigenstates, we show that the eigenstates are similar to the Jaynes-Cummings ladder of eigenstates and can be optically controlled by an external control field. Tuning the control field frequency to the one photon resonance, we show the superbunching behavior in the outgoing light field can be observed at the frequency of one photon resonance. We also show that there exists a magic control field intensity at which two photon blockade phenomenon can be significantly improved. In particular, it is possible to adjust the nonclassicality of the outgoing field from quantum to classical by varying the control field intensity. The work presented here provides an optical method to control statistical features of the outgoing field and can be useful for the nonclassical light generation, quantum gate operation and exotic quantum state generation.
\end{abstract}

\pacs{42.50.Pq, 42.50.Nn, 37.30.+i}

\maketitle

%
\section{Introduction}
Nonclassical properties of optical field such as squeezing, antibunching, and sub-Poissonian photon statistics, have been intensively studied in modern quantum optics~\cite{scully1997quantum,agarwal2012quantum}. Using these features of nonclassical optical field, many applications including the high-precision optical measurements, optical imaging, optical information processing and high-fidelity optical communications can be achieved based on the possibility of overcoming the so-called standard quantum limit~\cite{bachor2004guide,xiao1987precision,treps2002surpassing,grangier1987squeezed,polzik1992atomic,marin1997demonstration,yamamoto1990ii}. The generation of the nonclassical light still remains an open question, let alone a lot of attempts to effectively control the nonclassical optical fields.

In general, the field correlation function is an effective quantity to characterize the nonclassical optical field which is defined by $g^{(2)}(\tau)=\langle \hat{I}(t)\hat{I}(t+\tau)\rangle/\langle\hat{I}(t)\rangle^2$~\cite{brecha1999n,clemens2000nonclassical}. It reflects the probability of detecting one photon at time $t+\tau$ provided that one was detected at time $t$. According to the Schwartz inequality, the optical field is classical if the field correlation function $g^{(2)}(0)>g^{(2)}(\tau)$. The violation of this condition implies a nonclassical optical field such as antibunching [$g^{(2)}(0)<g^{(2)}(\tau)$] and sub-Poissonian distribution with $g^{(2)}(0)<1$.

In the past decade, many researches on the antibunching photons have been theoretically predicted and experimentally demonstrated in cavity quantum electrodynamics (QED) systems based on two photon blockade phenomenon~\cite{birnbaum2005photon}. In a single atom cavity QED system,  antibunching photons with sub-Poissonian [$g^{(2)}(0)<g^{(2)}(\tau)$ and $g^{(2)}(0)<1$] can be observed since $N>1$ photon transitions are blockaded due to the vacuum Rabi splitting. Up to date, the achievements of antibunching photons with sub-Poissonian have been reported in many configurations, including cavity QED system~\cite{dayan2008b,hamsen2017two,dayan2008b}, artificial atoms on a chip~\cite{Faraon2008Coherent,reinhard2012strongly}, cavity with Kerr nonlinearity~\cite{imamoglu1997strongly,rebic1999large,rebic2002photon} and superconducting circuits~\cite{lang2011aa,hoffman2011aj}.

Recently, a great deal of attention has been paid to the control of nonclassicality of the optical field. One of the most effective method is using the electromagnetically induced transparency (EIT) based on quantum interference effect~\cite{boller1991observation,fleischhauer2005electromagnetically,mucke2010electromagnetically}. By merging EIT configuration with the cavity QED system, the two photon blockade can be enhanced~\cite{rebic2002photon,souza2013coherent} and many interesting phenomenons have been theoretically proposed and experimentally demonstrated, including slow light propagation~\cite{zhang2008slow,nikoghosyan2010photon}, cavity cooling~\cite{kampschulte2010optical,bienert2012cavity,kampschulte2014electromagnetically}, cross phase modulation and quantum phase gate operation~\cite{zhu2010large,zou2013light,borges2016quantum}, all-optical switch and transistor~\cite{joshi2003optical,wei2010all,chen2013all,albert2011cavity,nielsen2011efficient} and quantum information processing~\cite{boozer2007reversible,himsworth2011eit}. In addition, cavity-assisted Rydberg-atom EIT phenomenon in a high-finesse optical cavity has been experimentally demonstrated~\cite{sheng2017intracavity}, and Y. Wu et al. explore the possibility of generating and controlling optical frequency combs in an cavity EIT system~\cite{li2018generation}.


In this work we present some interesting results for the electromagnetic control of the nonclassicality of the output field in an atom cavity QED system. Different from the early works on cavity EIT system~\cite{souza2013coherent,yang2015quantum}, we consider the case that the probe field directly drives the atom rather than the cavity. In addition, we tuned the frequency of the control field to be resonant with the one photon transition, which breaks the traditional EIT condition. By exploring the well-known ladder of eigenstates, we find that the eigenstates can be optically controlled by the external control field. We show that there exists a {\it magic} control field intensity where the two photon blockade can be significantly enhanced with reasonable photon number in the cavity, which provides a possibility to generate nonclassical light with antibunching behavior. At the frequency of one photon resonance, we show that the superbunching behavior may be observed if the control field is turned on. We also show that it is possible to optically control the nonclassicality of the outgoing light field by adjusting the external control field intensity, which provides an optical knob to tune the photon statistics of the outgoing light field.

\section{Theoretical model}
%
\begin{figure}[htb]
	\centering
	\includegraphics[width=8 cm]{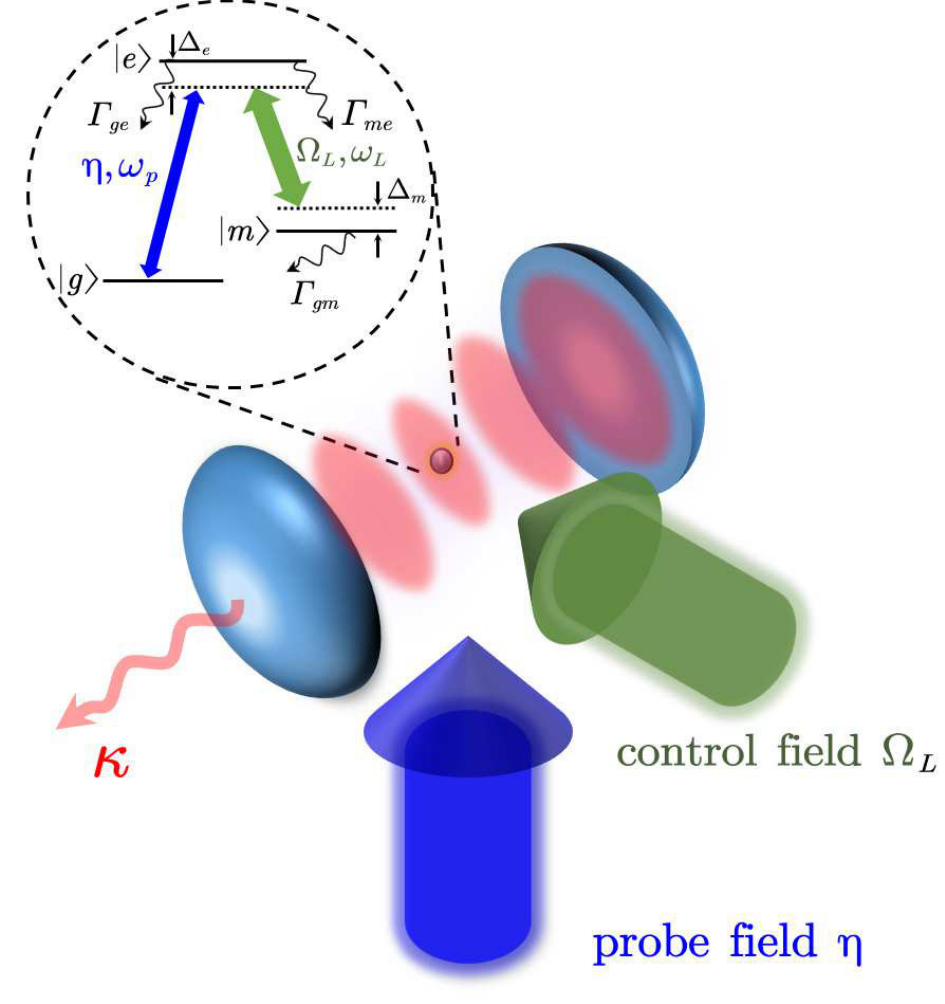}
	\caption{(Color online) The system configuration of a three-level $\Lambda$-type atom strongly coupled to a single mode cavity with wavelength $\lambda_{c}$. A probe field $\eta$ with angular frequency $\omega_p$ drives the $|g\rangle\leftrightarrow|e\rangle$ transition, and a control field $\Omega_L$ with angular frequency $\omega_{L}$ drives the $|m\rangle\leftrightarrow|e\rangle$ transition. Here, the cavity decay rate is denoted as $\kappa$. $\Gamma_{ge}$ and $\Gamma_{me}$ represent the spontaneous decay rates from $|e\rangle$ state to $|g\rangle$ and $|m\rangle$ state, respectively. The decay rate of the metastable state $|m\rangle$ is denoted as  $\Gamma_{gm}$. $\Delta_{e}=(\omega_{e}-\omega_g)-\omega_{p}$ and $\Delta_{m}=(\omega_{m}-\omega_g)-(\omega_p-\omega_{L})$ are the detunings of  $|e\rangle$ and $|m\rangle$ states, respectively.}
	\label{fig:fig1}
\end{figure}
To begin with, we consider a three-level $\Lambda$-type atom strongly coupled to a single-mode cavity with wavelength $\lambda_{c}$ (see Fig.~\ref{fig:fig1}). Here, we assume that the cavity mode frequency $\omega_c=2\pi/\lambda_c$ is equal to the resonant frequency between $|g\rangle$ and $|e\rangle$ states, i.e. $\omega_c=\omega_e-\omega_g$. A probe field with Rabi frequency $\eta$ couples the $|g\rangle\leftrightarrow|e\rangle$ transition, and a control field with Rabi frequency $\Omega_{L}$ couples the $|m\rangle\leftrightarrow|e\rangle$ transition.

Under the rotating-wave and electric dipole approximations, the Hamiltonian of this single atom cavity QED system can be written as
\begin{eqnarray}\label{eq:H}
 H&=&\Delta_{e}\sigma_{ee}+\Delta_{m}\sigma_{mm}+\Delta_{c}a^{\dagger}a+g(a\sigma_{eg}+a^\dagger\sigma_{ge}) \nonumber\\
 & &+(\Omega_{L}\sigma_{em}+\Omega^\dagger_{L}\sigma_{me})+\eta(\sigma_{eg}+\sigma_{ge})
\end{eqnarray}
where $a$ and $a^{\dag}$ are the annihilation and creation operators of the cavity field, and $\sigma_{ij}=|i\rangle\langle j|$ ($i,j=g,e,m$) are the atomic raising and lowering operators for $i\neq j$, and the atomic population operators for $i=j$. The detunings are defined as $\Delta_{c}=\omega_{c}-\omega_{p}$,  $\Delta_{e}=\Delta_c=(\omega_{e}-\omega_{g})-\omega_{p}$ and $\Delta_{m}=\Delta_p-\Delta_L$ with the control field detuning $\Delta_L=(\omega_{e}-\omega_{m})-\omega_{L}$. Here, the energy of $|j\rangle$ state is $\hbar\omega_j\ (j=g,e,m)$ and $\omega_{p(L)}$ is the angular frequency of the probe (control) field. In general, the properties of the entire system can be obtained by numerically solving the master equation, i.e.,
\begin{equation}\label{eq:master}
 \frac{d}{dt}\rho=-\frac{i}{\hbar}[H,\rho]+{\cal L}_{\rm atom}(\rho)+{\cal L}_{\rm cavity}(\rho)
\end{equation}
where $\rho$ is the density operator of the single atom cavity QED system, ${\cal L}_{\rm atom}(\rho)$ and ${\cal L}_{\rm cavity}(\rho)$ are the Liouvillian operators for the atomic decay and cavity decay, respectively, which are given by
\begin{eqnarray*}\label{eq:Latom}
{\cal L}_{\rm atom}(\rho)&=&\Gamma_{ge}(2\sigma_{eg}^{\dag}\rho \sigma_{eg}-\sigma_{eg}\sigma_{eg}^{\dag}\rho-\rho \sigma_{eg}\sigma_{eg}^{\dag})\\
&&+\Gamma_{me}(2\sigma_{em}^{\dag}\rho \sigma_{em}-\sigma_{em}\sigma_{em}^{\dag}\rho-\rho \sigma_{em}\sigma_{em}^{\dag})\\
&&+\Gamma_{gm}(2\sigma_{mg}^{\dag}\rho \sigma_{mg}-\sigma_{mg}\sigma_{eg}^{\dag}\rho-\rho \sigma_{mg}\sigma_{mg}^{\dag}),
\end{eqnarray*}
and
\begin{eqnarray*}\label{eq:Lcavity}
	{\cal L}_{\rm cavity}(\rho)&=&\kappa(2a \rho a^{\dag}-a^{\dag}a\rho-\rho aa^{\dag}),
\end{eqnarray*}
with $\Gamma_{ij}\ (i,j=g,e,m)$ being the decay rate from $|j\rangle$ to $|i\rangle$ state, and $\kappa$ being the decay rate of the cavity.

\section{eigenvalues and dressed state picture}
%
\begin{figure*}[htbp]
	\centering
	\includegraphics[width=17.0 cm]{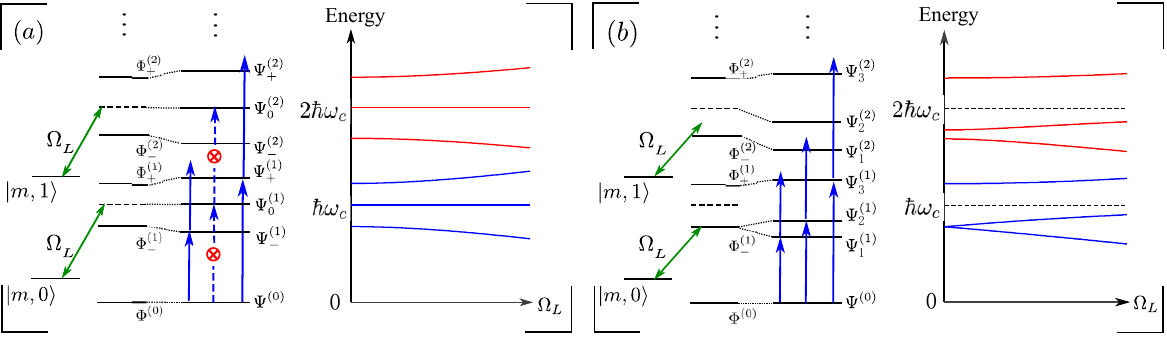}
	\caption{(Color online) The dressed states structure and the eigenvalues of this single atom cavity QED system with $\Delta_L=0$ [panel (a)] and $\Delta_L=-g$ [panel (b)], respectively. The blue solid arrow lines represent the allowed transitions for the probe field, but the dashed arrow lines represent the forbidden transitions. Here, the eigenvalues are obtained by numerically solving Eq. (3) with $g=20$.}
	\label{fig:fig2}
\end{figure*}
For any quantum system, it is helpful to study the eigenvalues and the corresponding eigenstates of the system, so that the physical mechanism of this system can be understand very well. Assuming $\eta=\Delta_{c}=\Delta_e=0$, the Hamiltonian can be rewritten in a new set of basis  $\{|g,n\rangle$,\ $|+,n-1\rangle$,\ $|-,n-1\rangle\}$ with $n$ being the number of photons in cavity and $|\pm,n-1\rangle=(|m,n-1\rangle\pm|e,n-1\rangle)/\sqrt{2}$. Therefore, in the $n$-photon space, the Hamiltonian can be expressed as~\cite{souza2013coherent}
\begin{equation}
H^{(n)}=\left(
\begin{array}{ccc}
0 & g\sqrt{n/2} & -g\sqrt{n/2}\\
g\sqrt{n/2} & \Omega_L+\Delta_L/2 & \Delta_L/2\\
-g\sqrt{n/2} & \Delta_L/2 & -\Omega_L+\Delta_L/2
\end{array}
\right).
\end{equation}

In the case of $\Delta_L=0$, the eigenvalues of the Hamiltonian $H^{(n)}$ can be solved analytically, yielding $\lambda_0^{(n)}=0$ and $\lambda_\pm^{(n)}=\pm\sqrt{\Omega_L^2+ng^2}$. The corresponding  eigenstates are given by
\begin{equation}
\Psi_0^{(n)}={\cal N}_0^{(n)}\left(|g,n\rangle-\frac{g\sqrt{n}}{2\Omega_L}|m,n-1\rangle\right),	
\end{equation}
and
\begin{eqnarray}
	\Psi_\pm^{(n)}&=&{\cal N}_\pm^{(n)}\left(|g,n\rangle+\frac{g\sqrt{n/2}}{\lambda_\pm^{(n)}-\Omega_L}|+,n-1\rangle\right.\nonumber\\
	& &\left.-\frac{g\sqrt{n/2}}{\lambda_\pm^{(n)}+\Omega_L}|-,n-1\rangle\right),
\end{eqnarray}
where ${\cal N}_0^{(n)}$ and ${\cal N}_\pm^{(n)}$ are the normalization factors. It is clear that the eigenstates $\Psi_n^{(0)}$ are the intracavity dark states which can not be excited. Then, the rest eigenstates form a ladder of energy levels which are arranged in doublets. The splitting between the doublets depends on the control field Rabi frequency $\Omega_L$ and the quantum number $n$ [see Fig. 2(a), right plot]. These results can also be explained by decomposing the system into two components. One is the subsystem consisting of the cavity and a two level atom with $|g\rangle$ and $|e\rangle$ states, and the other is the subsystem consisting of the control field and $|m\rangle$ state. As is known to all, the first subsystem has been studied in many literatures, which forms a ladder of doublet levels with energy splitting $2g\sqrt{n}$ [see Fig. 2(a), left plot]. The corresponding eigenvalues and eigenstates are given by $\Lambda_\pm^{(n)}=g\sqrt{n}$ and $\Phi_\pm^{(n)}=(\pm|g,n\rangle+|e,n-1\rangle)/\sqrt{2}$, respectively. Therefore, in the case of detuning $\Delta_L=0$, the energies of $\Phi_\pm^{(n)}$ states are shifted since the control field is far off resonant with each  state [see Fig. 2(a)].

Likewise, the control field can also be tuned resonant with the $|m,0\rangle\leftrightarrow|\Phi_-^{(1)}\rangle$ transition by choosing the detuning $\Delta_L=-g$. As a result, the $\Phi_-^{(1)}$ state is split into a doublet, but the $\Phi_+^{(1)}$ state undergoes an energy shift for off-resonant coupling [see Fig. 2(b), left plot]. Directing solving the eigenvalues of Hamiltonian given in Eq. (3), we show the eigenvalues with photon number $n=1,\ 2$ change with the control field Rabi frequency $\Omega_L$ in the left plot of Fig. 2(b). It is clear that the numerical results match very well with the analysis based on the dressed states. In the following, we will show that this energy splitting may result in a significant improvement of the nonclassicality of the cavity field if a magic control field intensity is applied. In addition, we will also show that the cavity field can be changed from nonclassical to classical by varying the control field intensity.

\section{the case of $\Delta_L=0$}
\begin{figure}[htbp]
	\centering
	\includegraphics[width=8.5 cm]{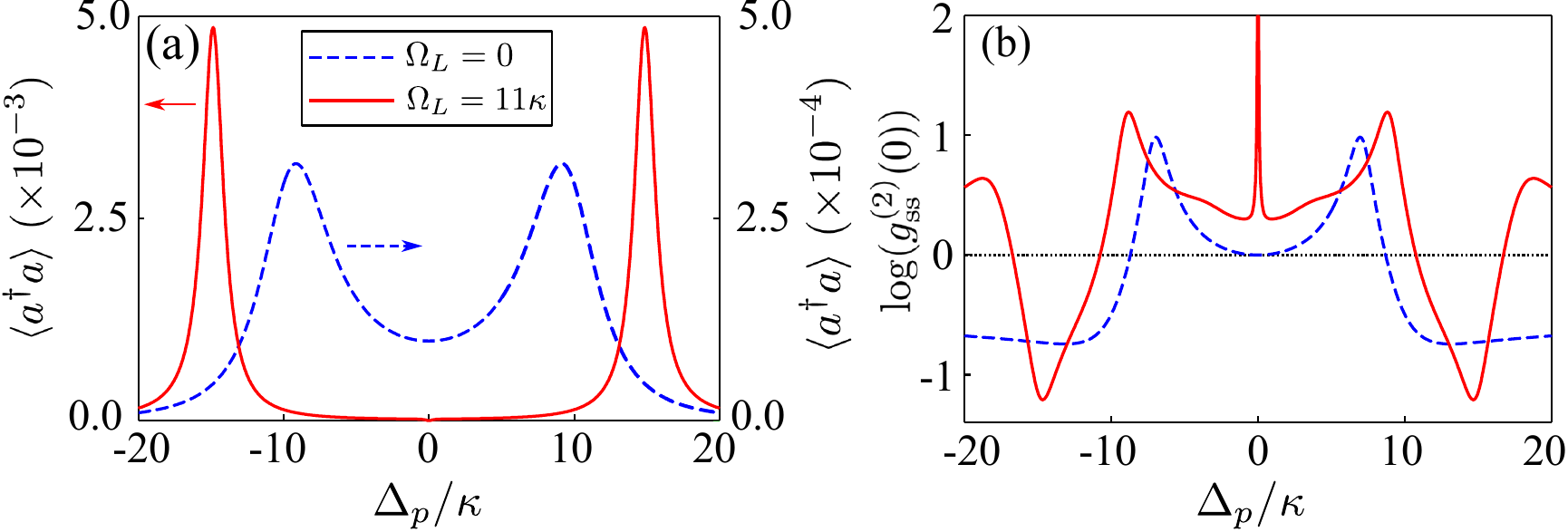}
	\caption{(Color online) Panels (a) and (b) show the mean photon number $\langle a^{\dag}a \rangle$ and the steady-state second-order photon-photon correlation function $\log{(g_{\rm ss}^{(2)}(0))}$ as a function of the normalized detuning $\Delta_p/\kappa$ for the probe field, respectively. The control field is taken as  $\Omega_{L}=0$ Hz (blue dashed curve) and $11.0\kappa$ (red solid curve), respectively. Other system parameters are chosen as $\Delta_e=\Delta_c=\Delta_L=0$ Hz, $\Gamma_{ge}=\Gamma_{me}=1.5\kappa$, $\Gamma_{gm}=5\times10^{-4}\kappa$, $g=10\kappa$ and  $\eta=0.1\kappa$. The dash-dotted line in panel (b) indicates the condition of $g^{(2)}_{\rm ss}(0)=1$.
	}
\label{fig:fig3}
\end{figure}

Before studying the nonclassicality of the cavity field, we first calculate the mean photon number $n_{\rm cav}=\langle a^\dag a\rangle$ in the cavity by numerically solving  Eq.~(\ref{eq:master}) without any approximation, which directly reflects the energy shifts of each eigenstate. In Fig.~\ref{fig:fig3}(a), we plot the mean photon number $n_{\rm cav}$ as a function of the normalized detuning $\Delta_p/\kappa$ for the probe field.
Here, we choose the control field Rabi frequency $\Omega_{L}/\kappa=0$ (blue dashed curve) and $11.0$ (red solid curve),  respectively. Other system parameters are chosen as $\Delta_e=\Delta_L=\Delta_c=0$ Hz, $\Gamma_{ge}=\Gamma_{me}=1.5\kappa$, $\Gamma_{gm}=5\times10^{-4}\kappa$, $g=10\kappa$ and $\eta/\kappa=0.1$. In the absence of the control field (i.e., $\Omega_L=0$ Hz), we can observe two peaks (see the blue dashed curve) at $\Delta_p=\pm g$ in the cavity excitation spectrum, which corresponds to two one-photon transitions, i.e., $|\Phi_0^{(0)}\rangle\rightarrow|\Phi_\pm^{(1)}\rangle$. In the presence of the control field, however, the position and amplitude of these two resonant peaks change greatly as the eigenvalues and eigenstates of the system are modified by the control field. As show in panel (a), we observe a larger energy splitting between two peaks (i.e., $2\sqrt{g^2+\Omega_L^2}$, see red solid curve), and the corresponding mean photon number also increases significantly. It is worth to notice that high order transitions are too weak to be observed since the multi ($n\geq2$) photon transitions are far off resonant, which results in the two photon blockade phenomenon.

To characterize this interesting phenomenon, we calculate the equal time second order photon-photon correlation function $g_{\rm ss}^{(2)}{(0)}=\langle a^\dag(0)a^\dag(0)a(0)a(0)\rangle/\langle a^\dag(0)a(0)\rangle^2$ at the steady state condition. As shown in Fig.~\ref{fig:fig3}(b), the second order correlation function $g^{(2)}_{\rm ss}{(0)}<1$ at the peaks in the cavity excitation spectrum (corresponding to the one photon transitions), which is the evidence of the two-photon blockade phenomenon. Another key feature of this system is the presence of strong photon bunching behavior ($g^{(2)}_{\rm ss}(0)\gg1$) at $\Delta_p=0$. In the absence of the control field, the probe field is off resonant with all states so that the quantum property of the cavity field is the same as that of the probe field, and the second order correlation function $g_{\rm ss}^{(2)}(0)=1$ (a coherent field). When the control field is turned on, additional states $\Psi_0^{(n)}$ appear and is resonant with the probe field, which results in a classical field generation with superbunching behavior, i.e., $g^{(2)}_{\rm ss}(0)\gg1$.

\begin{figure}[htbp]
	\centering
	\includegraphics[width=8.5 cm]{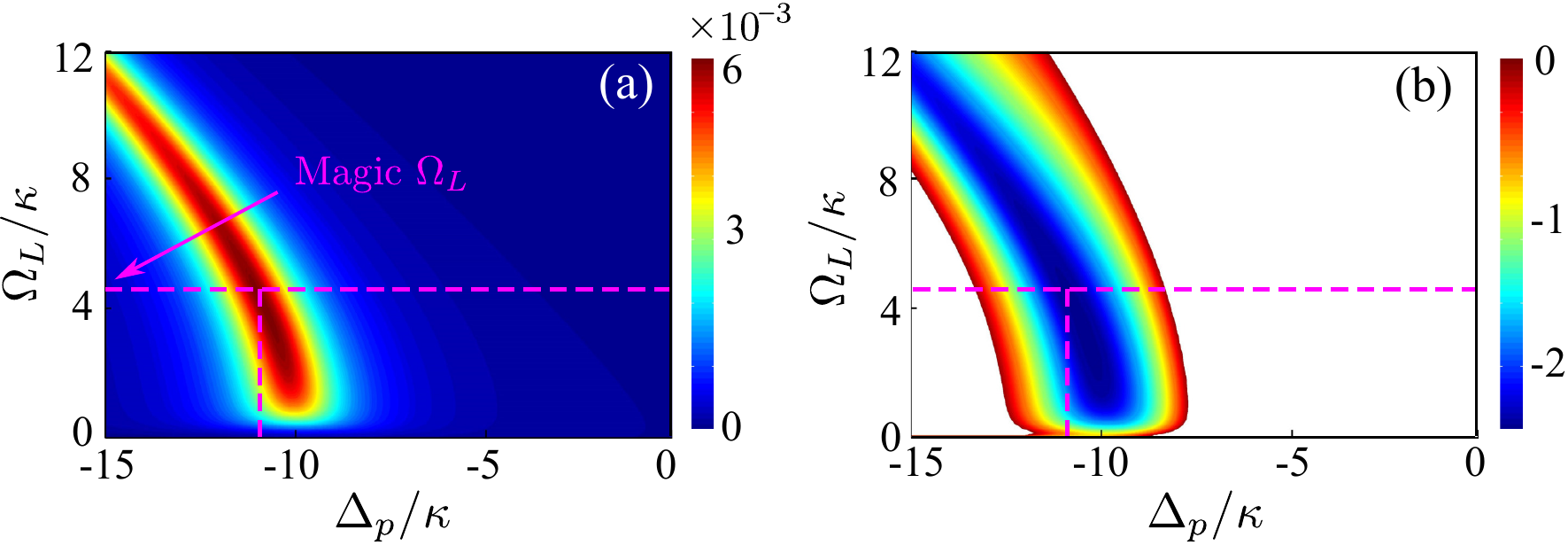}
	\caption{(Color online) Panels (a) and (b) show the mean photon number $n_{\rm cav}$ and steady state second order field correlation function $\log{(g^{(2)}_{\rm ss}(0))}$ as functions of the normalized detuning $\Delta_p/\kappa$ and control field Rabi frequency $\Omega_L/\kappa$, respectively. The pink horizontal dashed line indicates the magic control field Rabi frequency, where a specific probe field detuning (indicated by the pink vertical dashed line) can be chosen to realize an improved two-photon blockade phenomenon with reasonable photon number. The white area in panel (b) denotes $g_{\rm ss}^{(2)}(0)>1$.
	}
	\label{fig:fig4}
\end{figure}
In Fig.~\ref{fig:fig4}(a) and (b), we plot the mean photon number $n_{\rm cav}$ and the steady state second order field correlation function $\log{(g^{(2)}_{\rm ss}(0))}$ as functions of the normalized detuning $\Delta_p/\kappa$ and control field Rabi frequency $\Omega_L/\kappa$, respectively. Here, the system parameters are the same as those used in Fig.~\ref{fig:fig3}, and we just consider the case of blue detuning, i.e., $\Delta_p<0$ due to the symmetry of the system. As shown in panel (a), the width of the energy-level splitting almost increases linearly as the intensity of the control field enhances. Correspondingly, the second order field correlation function strongly depending on the cavity photon excitation varies significantly [see panel (b)]. In particular, there exists a {\it magic} control field intensity (indicated by the pink horizontal dashed line) to obtain the minimum of the second order field correlation function. As a result, an improved two photon blockade phenomenon with reasonable mean photon number can be achieved by choosing a specific probe field detuning indicated by the pink vertical dashed line. For example, at the detuning $\Delta_p/\kappa=-10.8$, one can obtain $g^{(2)}(0)\approx0.04$ and $n_{\rm cav}\approx0.006$  with $\Omega_L/\kappa=4.5$.

\section{the case of $\Delta_L=-g$}
\begin{figure}[htbp]
	\centering
	\includegraphics[width=8.5 cm]{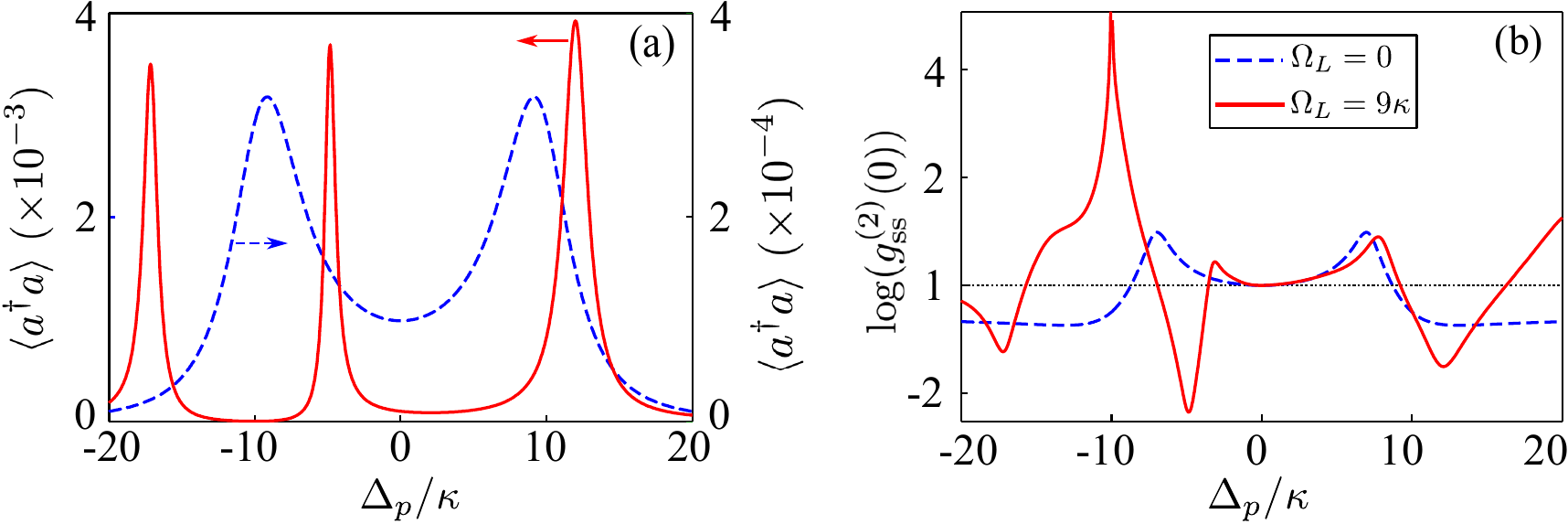}
	\caption{
		(Color online) The mean photon number $\langle a^{\dag}a \rangle$ and the steady-state second-order field correlation function $\log{(g_{\rm ss}^{(2)}(0))}$ are shown in panel (a) and (b), respectively. Here, we choose $\Delta_L=-g$ and the control field Rabi frequency is chosen as $\Omega_{L}/\kappa=0$ (blue dash curves) and $9.0$ (red solid curve), respectively. Other system parameters are the same as those used in Fig. 3. The dashed dotted line in panel (b) corresponds to $g_{\rm ss}^{(2)}(0)=1$.
	}
	\label{fig:fig5}
\end{figure}

Now, we consider the case of $\Delta_{L}=-g$, where the control field is tuned resonant with the $\Phi_{-}^{(1)}$ state shown in Fig. 2(b). In this case, the $\Phi_{-}^{(1)}$ state is split into two separate states labeled as $\Psi_1^{(1)}$ and $\Psi_2^{(1)}$, respectively. Because of the large detuning $\Delta_L$, the coupling between the control field and other states in the cavity QED system can be safely neglected. Therefore, there exist three resonant peaks in the cavity excitation spectrum as shown in Fig.~\ref{fig:fig5}(a), corresponding to the $|\Psi_0\rangle\leftrightarrow|\Psi_1^{(1)}\rangle$, $|\Psi_0\rangle\leftrightarrow|\Psi_1^{(2)}\rangle$ and $|\Psi_0\rangle\leftrightarrow|\Psi_1^{(3)}\rangle$ transitions, respectively.
Here, the control field is taken as $\Omega_{L}/\kappa=9.0$ (red solid curve), and other system parameters are the same as those used in Fig.~\ref{fig:fig3}. It is noted that the mean photon number in the cavity is significantly enhanced by the control field compared with the case of $\Omega_L=0$. In panel (b), we plot the steady state second order field correlation function $g^{(2)}_{\rm ss}(0)$ as a function of the normalized detuning $\Delta_p/\kappa$ with the control field Rabi frequency $\Omega_L/\kappa=0$ and $9$, respectively. It is clear that, in the presence of the control field, the superbunching behavior (i.e., $g^{(2)}_{\rm ss}(0)\gg1$) can be observed at $\Delta_p=-g$, where $g^{(2)}_{\rm ss}(0)<1$ if the control field is turned off. In addition, the second order field correlation function drops quickly at the frequency near the middle peak in the cavity excitation spectrum, which provides possibilities to achieve a significant improvement of the two photon blockade phenomenon.

\begin{figure}[htbp]
	\centering
	\includegraphics[width=8.5 cm]{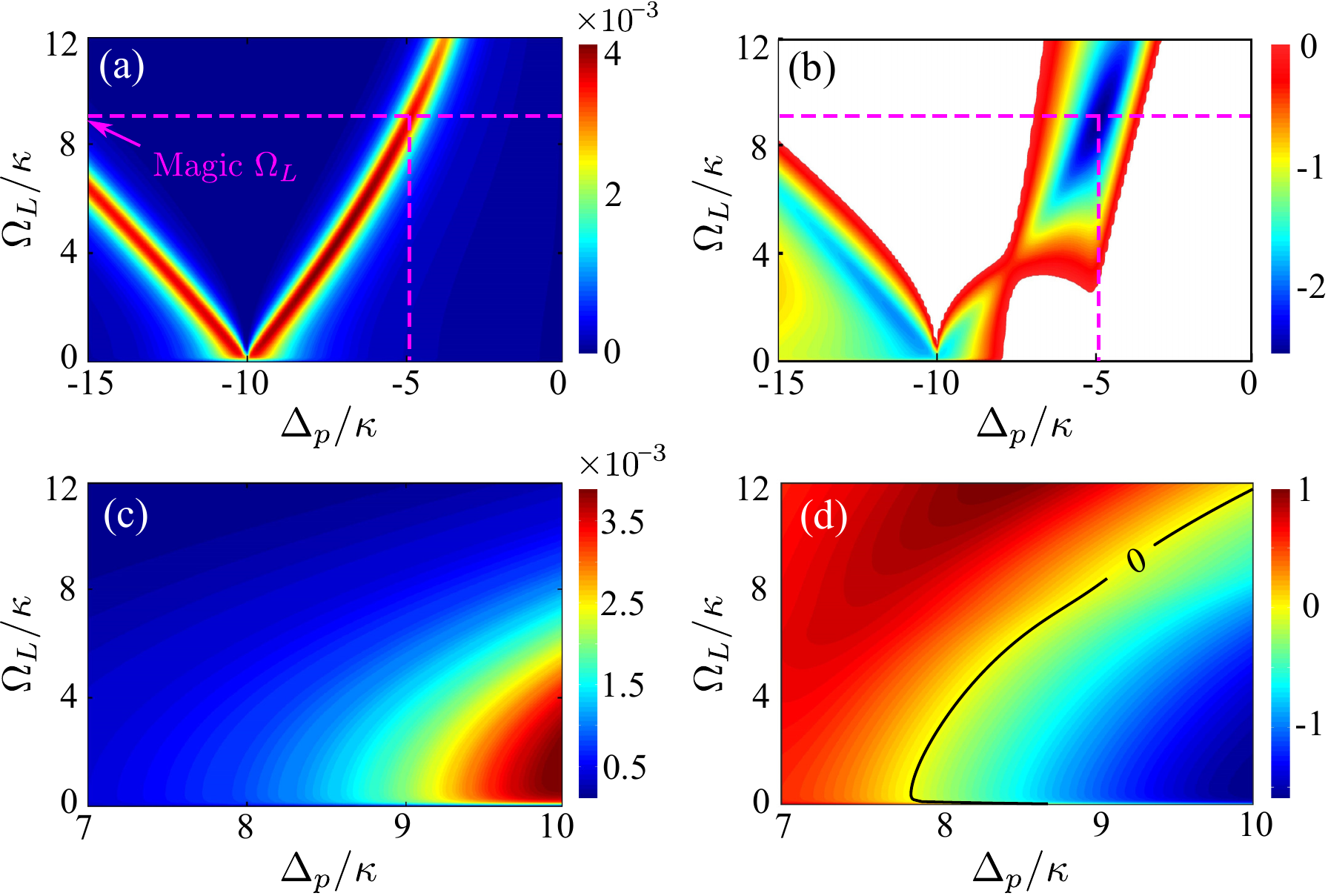}
	\caption{
		(Color online) Panels (a) and (c) show the mean photon number $n_{\rm cav}$, and panels (b) and (d) show the steady state second order field correlation function $\log{(g^{(2)}_{\rm ss}(0))}$. The pink horizontal dashed line in panels (a) and (b) indicates the magic control field Rabi frequency, where a specific probe field detuning (indicated by the pink vertical dashed line) can be chosen to achieve a significantly improved two-photon blockade phenomenon. The white areas in panel (b) denote the regime with $g_{\rm ss}^{(2)}(0)>1$. The black curve in panel (d) indicate the equal attitude line of $g_{\rm ss}^{(2)}(0)=1$.
	}\label{fig:fig6}
\end{figure}
Figure~\ref{fig:fig6}(a) and (b) show the mean photon number $n_{\rm cav}$ and the steady state second order field correlation function $g_{\rm ss}^{(2)}(0)$ as functions of the normalized control field Rabi frequency $\Omega_L/\kappa$ and detuning $\Delta_L/\kappa$, which is chosen near the middle peak. As shown in panel (a), the width of the energy splitting induced by the control field is almost proportional to the control field Rabi frequency. In particular, we can obtain a {\it magic} control field intensity indicated by the pink horizontal dashed line, where the two photon blockade phenomenon can be significantly improved when a  specific probe field detuning is chosen (indicated by the vertical line). According to our numerical calculation, the mean photon number $n_{\rm cav}\approx0.004$ and the second order field correlation function $g_{\rm ss}^{(2)}(0)\approx0.004$ at $\Delta_p/\kappa=-4.8$. Therefore, a nonclassical field with antibunching behavior is generated in this three level atom cavity QED system via the electromagnetic control. In Fig.~\ref{fig:fig6}(c) and (d), we choose the detuning $\Delta_L$ near the right peak in the cavity excitation spectrum. As shown in panel (d), the second order field correlation function changes from antibunching to superbunching by increasing the control field intensity. Correspondingly, the property of the output field can be controlled from quantum to classical by adjusting the control field.

\section{Conclusion}
In conclusion, we have shown that the quantum properties of the output field in a three level atom cavity QED system can be actively contrlled by an electromagnetic field. This arises from the dressed state structure of the system formed by the interacting fields and atom. The dressed state structure and allowed transitions
strongly depend on the control field intensity and frequency. We show that the energy splitting and photon blockade can be enhanced in the case of the control field detuning $\Delta_L=0$. In addition, the superbunching behavior of the output cavity field can be observed in this atom cavity QED system. In the case of $\Delta_L=-g$, we show that the significantly improved two photon blockade phenomenon can be observed if a magic control field intensity is chosen. We also show that the property of the outgoing light field can be controlled from quantum to classical by increasing the control field intensity. All these features can be explained by exploring the dressed state structure adjusted by the control field, and will be useful for the nonclassical light generation, optical controlled quantum gate operation and exotic quantum state preparation.

\begin{acknowledgments}
We thank G. S. Agarwal in Texas A\&M university for stimulating discussions. We acknowledge the National Key Basic Research Special Foundation (Grant No. 2016YFA0302800); the Joint Fund of the National Natural Science Foundation (Grant No. U1330203); the National Nature Science Foundation (Grant No. 11504272, 11774262, 11474003, 11504003); the Shanghai Science and Technology Committee (STCSM) (Grants No. 15XD1503700); the Anhui Provincial Natural Science Foundation (Grant No. 1408085MA19, 1608085ME102).
\end{acknowledgments}

\bibliographystyle{apsrev4-1}
\bibliography{references}

%
%
\end{document}